# DIGITAL SIGNAL TRANSMISSION WITH CHAOTIC ENCRYPTION: DESIGN AND EVALUATION ON FPGAS

Claudio M. González, Hilda A. Larrondo, *Member of IEEE*, Carlos A. Gayoso, Leonardo J. Arnone and Eduardo I. Boemo

**ABSTRACT.**

**A discrete-time discrete-value pseudo-chaotic encoder/decoder system is presented. The pseudo-chaotic module is a 3D discrete version of the well-known Lorenz dynamical system. Scaling and biasing transformations as well as natural number arithmetics are employed in order to simplify realizations on a small size Field Programmable Gate Array (FPGA). The encryption ability is improved by using only the least significant byte of one of the pseudo chaotic state variables as the key to encrypt the plain text. The key is periodically perturbed by another chaotic state variable. The statistical properties of the pseudo chaotic cipher are compared with those of other pseudo-random generators available in the literature. As an example of applicability of the technique, a full duplex communication system is designed and constructed using FPGA as technological framework.**

# I. Introduction

During the last years the use of chaos in communications has become a very important topic of research. Both analog and digital schemes have been proposed [1-7]. For a complete review on the state of the art, see [8-12].

Chaotic digital communications schemes may be coarse classified in one of two categories: 1) *chaotic modulation techniques* and 2) *chaotic encryption techniques.* In the first class, a continuous chaotic signal is modulated by the information and the emerging compound signal is digitized and transmitted through the communication channel. In the case of chaotic encryption techniques, a sequence generated by a discrete or continuous chaotic system is used as a *key* for ciphering a *plain text*.

Chaotic masking, switching, parameter modulation and frequency hopping sequences as well as symbolic representation of controlled chaotic orbits, are some of the methods proposed in the literature to send binary messages.

The level of security of analog and digital chaotic communication systems is an issue under discussion. Several unmasking procedures and also new realizations with enhanced security have recently appeared [13-22].

This paper contributes to the above lines by presenting a discrete-time discrete-value pseudo-chaotic cipher that employs one of the state variables of a Lorenz 3D-map. The main differences with previous schemes are: a) instead of a 1D map, a 3D map is used as *key sequence generator*; b) only the least significant byte of the pseudo-chaotic state variable is used in the encryption process. In addition, it is periodically perturbed by another pseudo-chaotic state variable of the own system, in order to improve the statistical properties; c) instead of floating point numbers (required by continuous-value ciphers), natural numbers are used here to simplify the hardware realization. As the system is asynchronous the Byte Error

Ratio (BER) present a high sensitivity respect to any minimal mismatch between receiver and transmitter. The use of natural number arithmetics simplifies this strict matching requirement; d) The full duplex communications system may be realized on 54K equivalent gate FPGA.

The organization of this paper is as follows: in section II, the chaotic encoder is presented and its statistical properties are analyzed; in section III, a complete communication system is prototyped using FPGA. Area-time figure of the sistem is evaluated in section IV.

## II. THE LORENZ 3D MAP: STATISTICAL PERFORMANCE AND REALIZATION

The Lorenz chaotic attractor is generated by the following continuous dynamical system:

$$\begin{cases} dx/dt = -\delta(x-y) \\ dy/dt = -xz + \Gamma x - y \\ dz/dt = xy - bz \end{cases} , \tag{1}$$

where δ, Γ and *b* are control parameters.

This continuous system may be converted into a 3D-map by means of the Euler´s first order approach, as follows:

$$\begin{cases} \tilde{x}_{n+1} = \tilde{x}_n + k[-\delta(\tilde{x}_n - \tilde{y}_n)] \\ \tilde{y}_{n+1} = \tilde{y}_n + k[-\tilde{x}_n\tilde{z}_n + \Gamma\tilde{x}_n - \tilde{y}_n] \\ \tilde{z}_{n+1} = \tilde{z}_n + k[\tilde{x}_n\tilde{y}_n - b\tilde{z}_n] \end{cases} , \tag{2}$$

where *k* is a time scaling parameter. In order to reduce the hardware complexity the following biasing and scaling transformations are applied:

$$\begin{cases} x_n = (\tilde{x}_n + B)S \\ y_n = (\tilde{y}_n + B)S \\ z_n = (\tilde{z}_n + B)S \end{cases} , \tag{3}$$

where *B* and *S* are respectively the biasing and scaling parameters. The resulting equations are:

$$
\begin{cases}
x_{n+1} = x_n + k\delta(y_n - x_n) \\
y_{n+1} = (1-k)y_n + (kB + k\Gamma)x_n + kBz_n - k/S\, x_n z_n \\
\qquad\quad + (kBS - k\Gamma BS - kB^2 S) \\
z_{n+1} = (1-kb)z_n - kB(x_n + y_n) + k/S\, x_n y_n \\
\qquad\quad + (kB^2 S + kbBS)
\end{cases}
\tag{4}
$$

In this paper, the following values are adopted:

$$
k = 1/64;\ \delta = 8;\ \Gamma = 24;\ b = 2;\ B = 40;\ S = 512 \tag{5}
$$

Thus, the final equations are:

$$
\begin{cases}
X_{n+1} = X_n + Y_n/8 - X_n/8 \\
Y_{n+1} = Y_n - Y_n/64 + X_n + Z_n/2 + Z_n/8 \\
\qquad\quad - (X_n/256)(Z_n/128) - 20160 \\
Z_{n+1} = Z_n - Z_n/32 + X_n - (X_n + Y_n)/2 \\
\qquad\quad + (X_n + Y_n)/8 \\
\qquad\qquad + (X_n/256)(Y_n/128) + 13440
\end{cases}
\tag{6}
$$

The map (6) is structurally stable. The operations involved are additions, subtractions, divisions by powers of two, and finally only two products.

Fig. 1 is a 3D-view of a trajectory of map (6) starting at $X_0$=18503, $Y_0$=21315 and $Z_0$=32032. The typical butterfly shape of the Lorenz attractor clearly emerges from this figure. The range of the state variables in Fig. 1 implies that at least 17 digits are required. The most significant bits of the state variables carry low frequency information concerning the Lorenz attractor, as will be shown below. To make the encryption process stronger, these bits must not be send through the channel.

It is important to note that the map (6) is not truly chaotic because the number of digits is finite, and the binary divisions produce a truncation effect. However, by increasing the scale factor S, it is possible to increase the period of the generated sequence because the truncation effect diminishes. Numerical experiments give up 78,782 as the highest attainable value. This period would be not high enough for specific applications. The use of small pseudo-chaotic perturbations increases this value to 6,500,000, if the perturbation is applied each 10,000 clock cycles.

Fig. 2 shows the block diagram of the 3D-map generator. The least significant byte of $X_n$ is the *pseudo-chaotic key,* key[7..0] used to encrypt the *plain text*. It is perturbed each N clock cycle, performing a XOR operation with the least significant byte of $Y_n$, producing the signal $X'_n$[7..0]. The highest bits of $X'_n$, it means $X'_n$[16..8] are identical to $X_n$[16..8]. $X'_n$[16..0] is reinjected in the dynamical system.

The cipher generated by the Lorenz 3D-map of Fig. 2 was tested using five standard statistical benchmarks. Table I shows a comparison between the technique presented in this paper and those obtained for three pseudo-random sequence generators reported in the literature [23, 24]. We took five samples for each generator. Table I shows that the Lorenz 3D-map passes all tests

Using this approach, the communication system of Fig. 3 has been designed as a case-study. Fig. 3a shows the block diagram of the transmitter. The *plain text* is the one byte length binary sequence labeled plain_text[7..0]; it means that 256 levels are used to obtain it by a digital conversion of the message. In the transmitter, the key[7..0] is mixed by means of a XOR operation with each byte of the plain text. The output of the XOR, labeled, ciper_text[7..0], leaves the transmitter after a parallel-serial conversion. It travels through the communication channel and enters into the receiver (see Fig. 3b). In the receiver a similar Lorenz 3D-map, starting from identical initial conditions, is used to reverse the process. The signal rec_text[7..0] is the recovered plain text.

## III. FPGA IMPLEMENTATION OF THE FULL COMMUNICATION SYSTEM

The Lorenz 3D-map was implemented on the Altera EPF10K20TC144-3 FPGA sample, using the standard Max Plus II designing tool. An amount of 676 Logic Cells (LC's) are required (i.e. less than the 58% of the logical resources available). The dynamics of the 3D-map is determined by the initial conditions ($X_{in}$[16..0], $Y_{in}$[16..0], $Z_{in}$[16..0]), and the repetition rate of the chaotic perturbation (N[13..0]).

The full transmitter (Fig. 3a) requires 13,800 gates and the full receiver (Fig 3b) 13,200 gates: that is, the 69% and 66% of one chip resources respectively

Figs. 4a and 4b show the timing diagram in the case of a plain text consisting of a repetitive sequence of the numbers 0 to 255 (see also Fig. 3). The transmitter and receiver clocks, that control the overall encryption/decryption process, run at 6.25 MHz and 25 MHz respectively. The transmission rate is 568.2Kbytes/s. As can be seen the recovered text , rec_text[7..0], is identical to the plain_text[7..0].

## IV. Evaluation of the full system

Several tools were utilized to evaluate the statistical properties of the encrypted signal. Figs. 5a and 5b evidence that the first and second order distributions of the encrypted signal, cipher_text[7...0], are remarkably uniform. The Fast Fourier Transform (FFT), (see Fig. 6a) looks like white noise. The inclusion of the most significant bits of Xn reinforces the low frequency spectrum as is depicted in Fig. 6b. Fig. 7 has four panels showing the importance of discarding the 9 most significant bits of $X_n$ , and the effect of the chaotic perturbation generated by the XOR gate in the block diagram of Fig. 2. These four panels show the autocovariance of the ciphered text $C_n$ in the following cases: in 7a (corresponding to the block diagram of Fig. 2) the key consists of only the 8 least significant bits of $X_n$ and the chaotic perturbation is "on". In (7b), the key consists of only the 8 least significant bits of $X_n$ but the chaotic perturbation is "off". In the third panel (7c) the key is $X_n$[16..0] and the chaotic perturbation is "on". Finally in the fourth panel (7d) the key is $X_n$[16..0] and the chaotic perturbation is “off”. These pictures indicate that the best choice is (a), i.e. the realization proposed in this paper.

In the case of an ideal channel considered here, the Bit Error Ratio (BER) must be 0 for identical Lorenz Generators at both ends and it must increase if synchronization is lost. In Fig. 8a the case when parameters are not identical is considered (parameter mismatching). The picture shows the BER as a function of the percentage of parameter mismatch between transmitter and receiver. In Fig. 8b the case of initial condition mismatching. It is clear that the system is very sensitive to parameter or initial condition mismatchings. Even the smallest possible didfference (the least significant bit) between one parameter or initial contition in the receiver, with the corresponding parameter or initial condition in the transmitter, leads to a complete desynchronization, and the BER grows up to its maximal value 0.5. It means that only a matched receiver can recover the plain text. Let us stress that the perfect matching is attainable in this digital system employing natural numbers. This is an important advantage over analog chaotic communication systems based on identical synchronization.

## V CONCLUSIONS

In this work, a digital communication system, using a pseudo-chaotic attractor as encrypting key, is presented. Natural number arithmetics has been employed in order to simplify the realization on 54K equivalent gate FPGA. Additionally, it makes straightforward the matching process between transmitter and receiver. A simple perturbation procedure used to increase the period of the pseudo-chaotic key, is proposed. The perturbation is generated by another pseudo chaotic variable of the same dynamical system. The statistical tests of the key, and the FFT and autocovariance of the ciphered text, comfirm that the encryption ability is enhanced by using the key proposed here: the least significant byte of the state variable, periodically perturbed by another pseudo-chaotic state variable. The high sensitivity of the BER to any minimal difference between transmitter and receiver, assures that only a matched receiver can decode the message. The functional and timing simulations and the compilation and fitting resources, show that the full duplex system may be implemented on four small FPGA.

## ACKNOWLEDGMENTS

The authors want to thank Dr. Eloy Anguiano for his suggestion of using the least significant bits of the pseudo-chaotic signals.

## FIRST PAGE FOOTNOTES

Manuscript received ___________________.

This work was partially supported by CONICET (PIP 330/98), ANPCyT and UNMDP (PICTO 11-090076) and the Spanish Ministry of Science and Technology (under Contract TIC2001-2688-C03-03). Claudio Marcelo González, Hilda Ángela Larrondo, Carlos Arturo Gayoso and Leonardo José Arnone are in Facultad de Ingeniería, Universidad Nacional de Mar del Plata. Juan B. Justo 4302, C. P. 7600, Mar del Plata – Argentina (larrondo@fi.mdp.edu.ar)

Eduardo Boemo is in the School Engineering - Universidad Autónoma de Madrid, Ctra. de Colmenar Km. 15, 28049 Madrid – Spain (eduardo.boemo@uam.es)

The corresponding author (Hilda Angela Larrondo) is also a CONICET researcher.

## FIGURE CAPTIONS

**Fig. 1.** 3D-view of the trajectory of the Lorenz 3D-map of eq. (6) starting at $X_0$=18503, $Y_0$=21315 and $Z_0$=32032.

**Fig. 2.** Block diagram of the dynamical system generating the 3D-map of eq. (6)

**Fig. 3.** (a) Block diagram of the transmitter; (b) Block diagram of the receiver

**Fig. 4.** Timing simulation performed with Max Plus II on the EPF10K20TC144-3 IC. (a) main signals involved in the transmitter (see also Fig. 3a); (b) main signals in the receiver (see also Fig. 3b).

**Fig. 5.** Trajectory plots for $C_n$=out_rx in the case of a plain text consisting in a periodic sequence of the numbers 0 to 255. (a) 2D trajectory plot for $C_n$=out_rx; (b) 3D trajectory plot for $C_n$=out_rx.

**Fig. 6.** Fast Fourier Transform (FFT) of the ciphered text $C_n$. The plain text is a periodic sequence of the numbers 0 to 255. The FFT is evaluated over 100,000 samples. (a) only the least significant byte of $X_n$ are used in the encryption process. (b) all the 17 bits of $X_n$ are used in the encryption process.

**Fig. 7.** Autocovariance of the ciphered text $C_n$ evaluated over 1,000,000 samples: (a) only the least significant byte of $X_n$ is used in the encryption process and perturbation is on; (b) only the least significant byte of $X_n$ is used in the encryption process and perturbation is off; (c) all the 17 bits of $X_n$ are used in the encryption process and perturbation is on; (d) all the 17 bits of $X_n$ are used in the encryption process and perturbation is off.

**Fig. 8.** Relationship between BER and percentage of receiver mismatches. (a) BER versus parameter $p=(kB^2S+kbBS)$ of equation 4. (b) BER versus a mismatch in the initial condition $X_0$

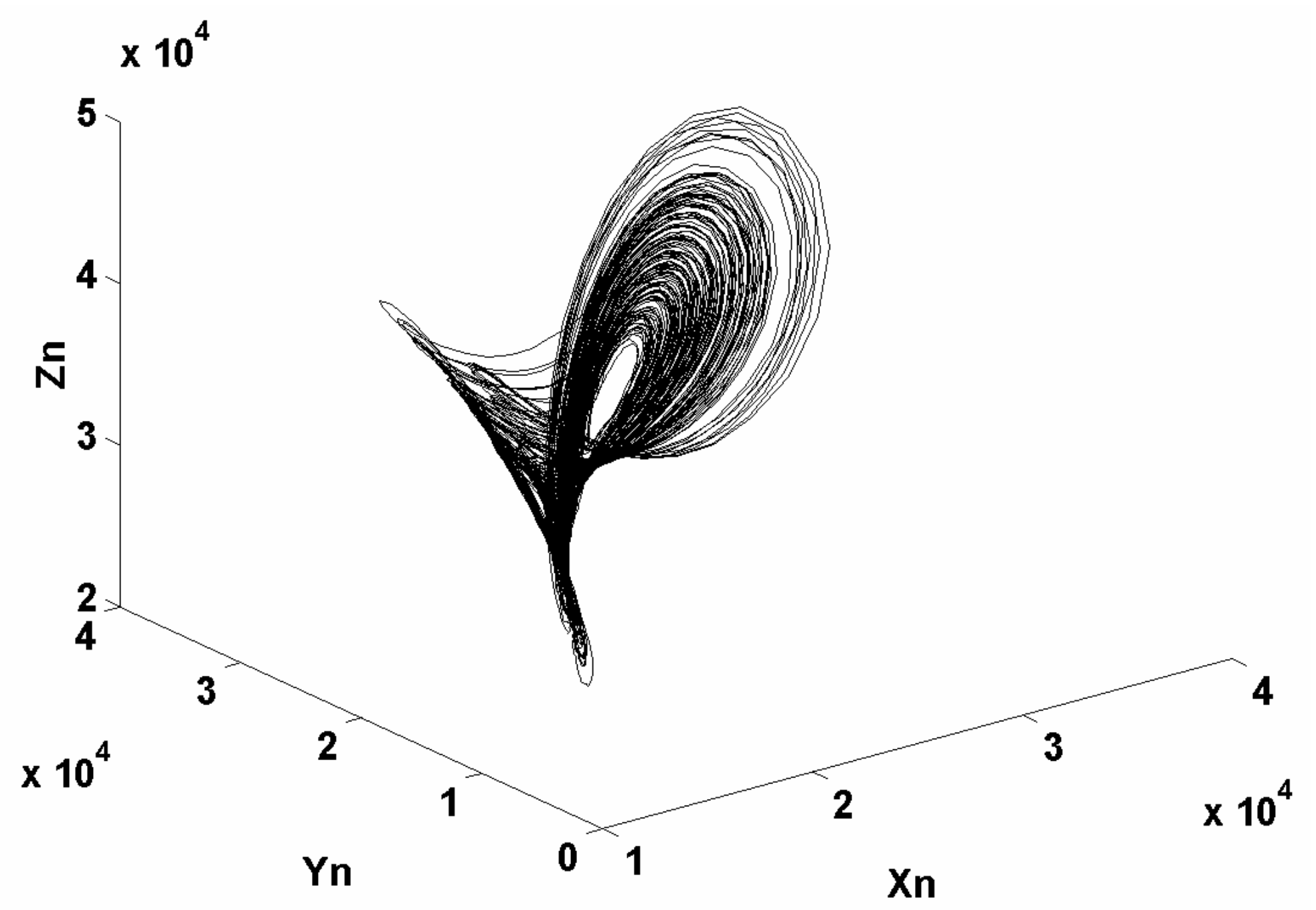

**Fig. 1.** 3D-view of the trajectory of the Lorenz 3D-map of eq. (6) starting at $X_0$=18503, $Y_0$=21315 and $Z_0$=32032.

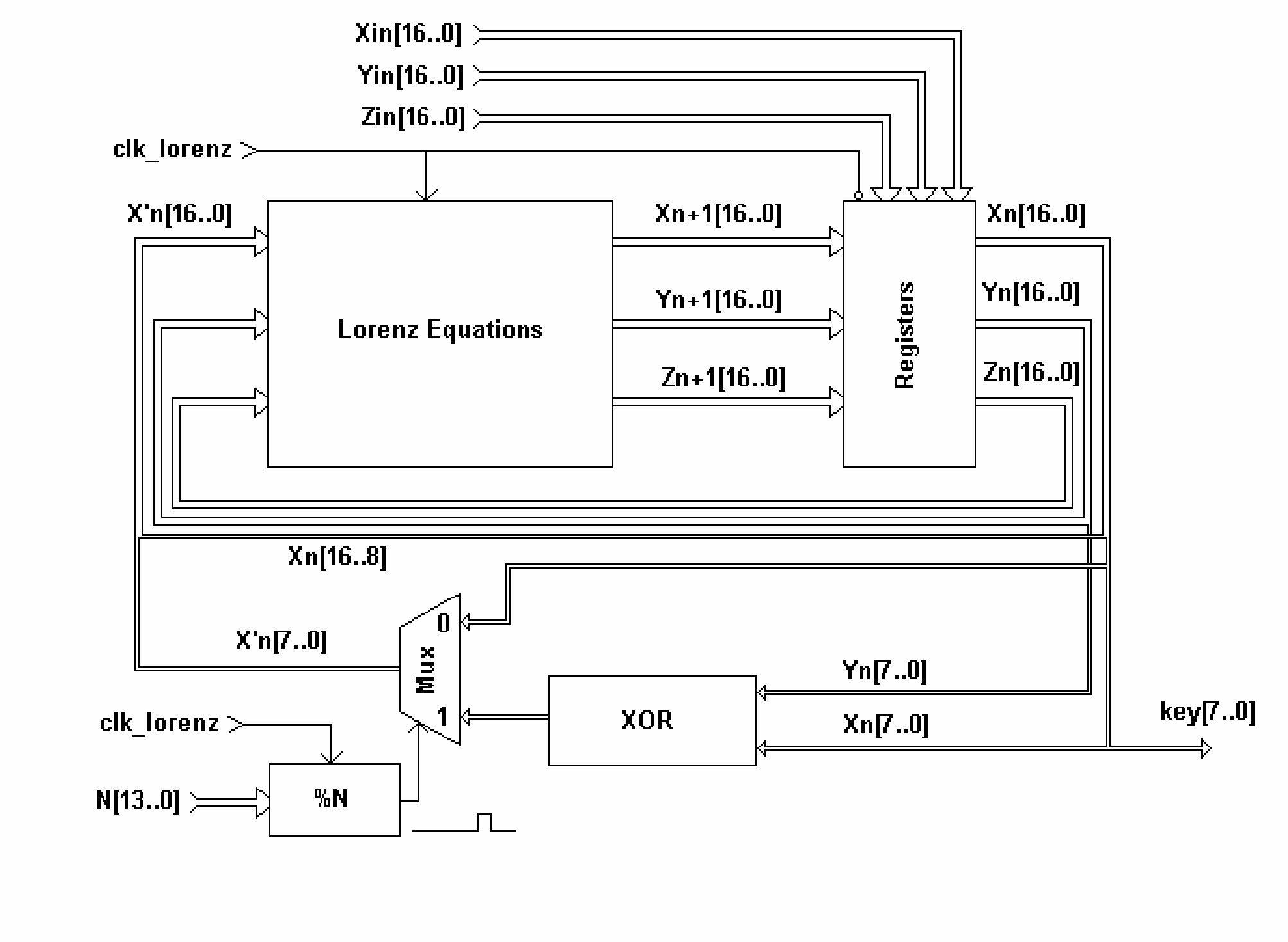

**Fig. 2.** Block diagram of the dynamical system generating the 3D-map of eq. (6)

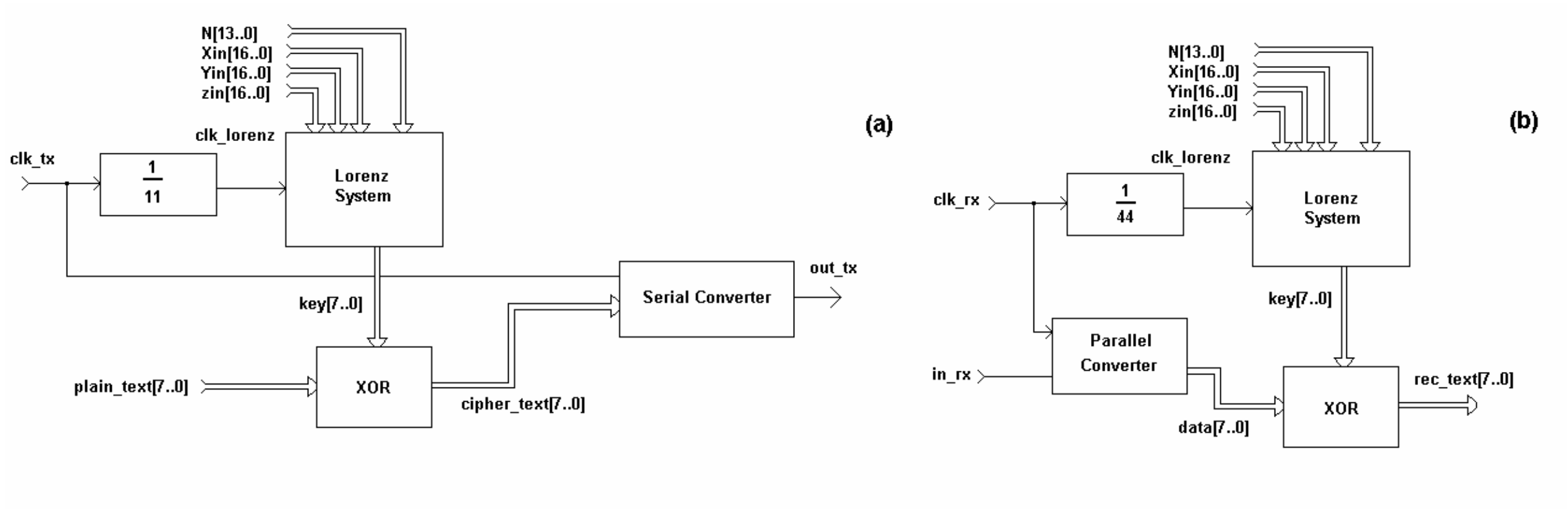

**Fig. 3.** (a) Block diagram of the transmitter; (b) Block diagram of the receiver

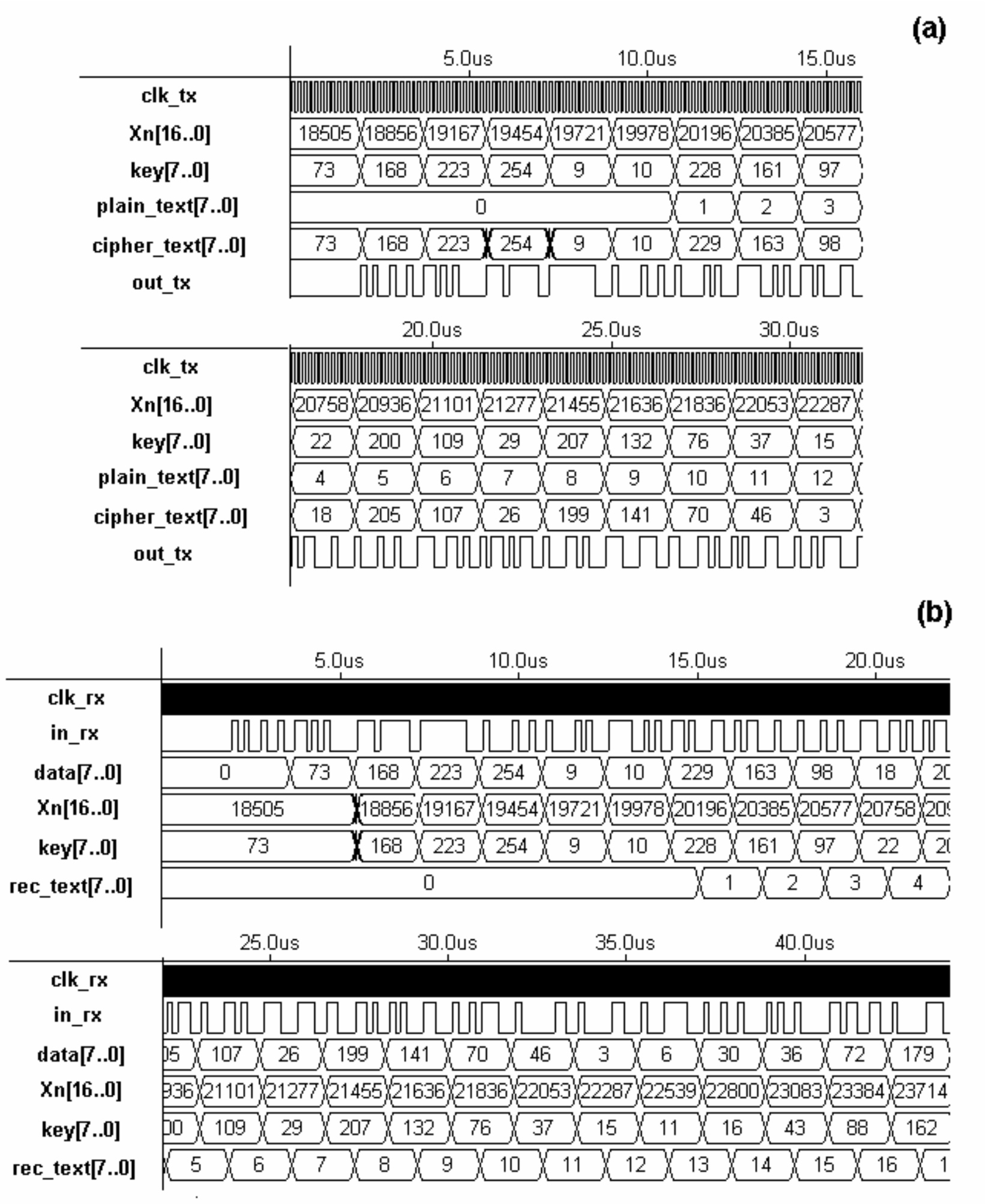

**Fig. 4.** Timing simulation performed with Max Plus II on the EPF10K20TC144-3 IC. (a) main signals involved in the transmitter (see also Fig. 3a); (b) main signals in the receiver (see also Fig. 3b).

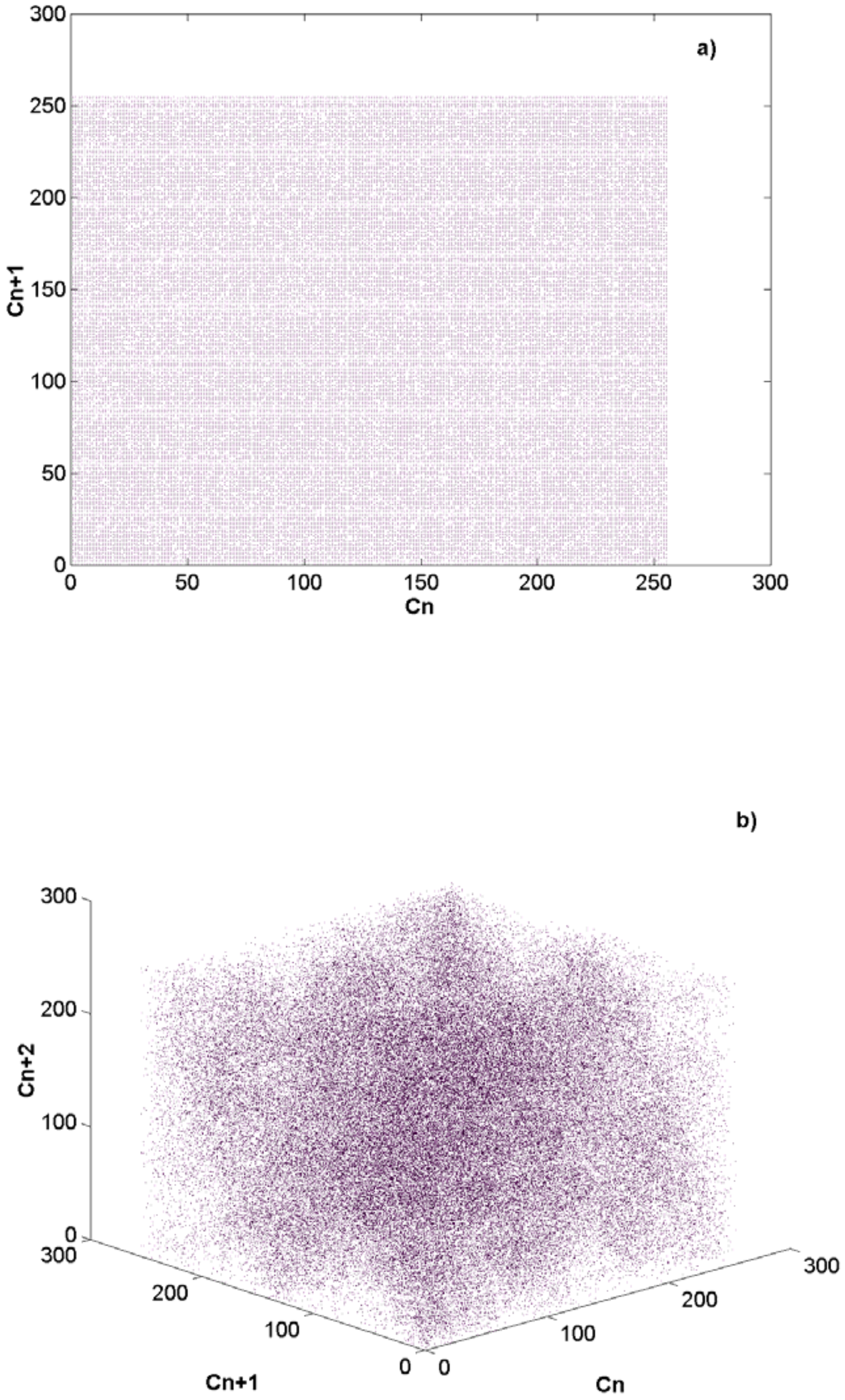

**Fig. 5.** Trajectory plots for $C_n$=out_rx in the case of a plain text consisting in a periodic sequence of the numbers 0 to 255. (a) 2D trajectory plot for $C_n$=out_rx; (b) 3D trajectory plot for $C_n$=out_rx.

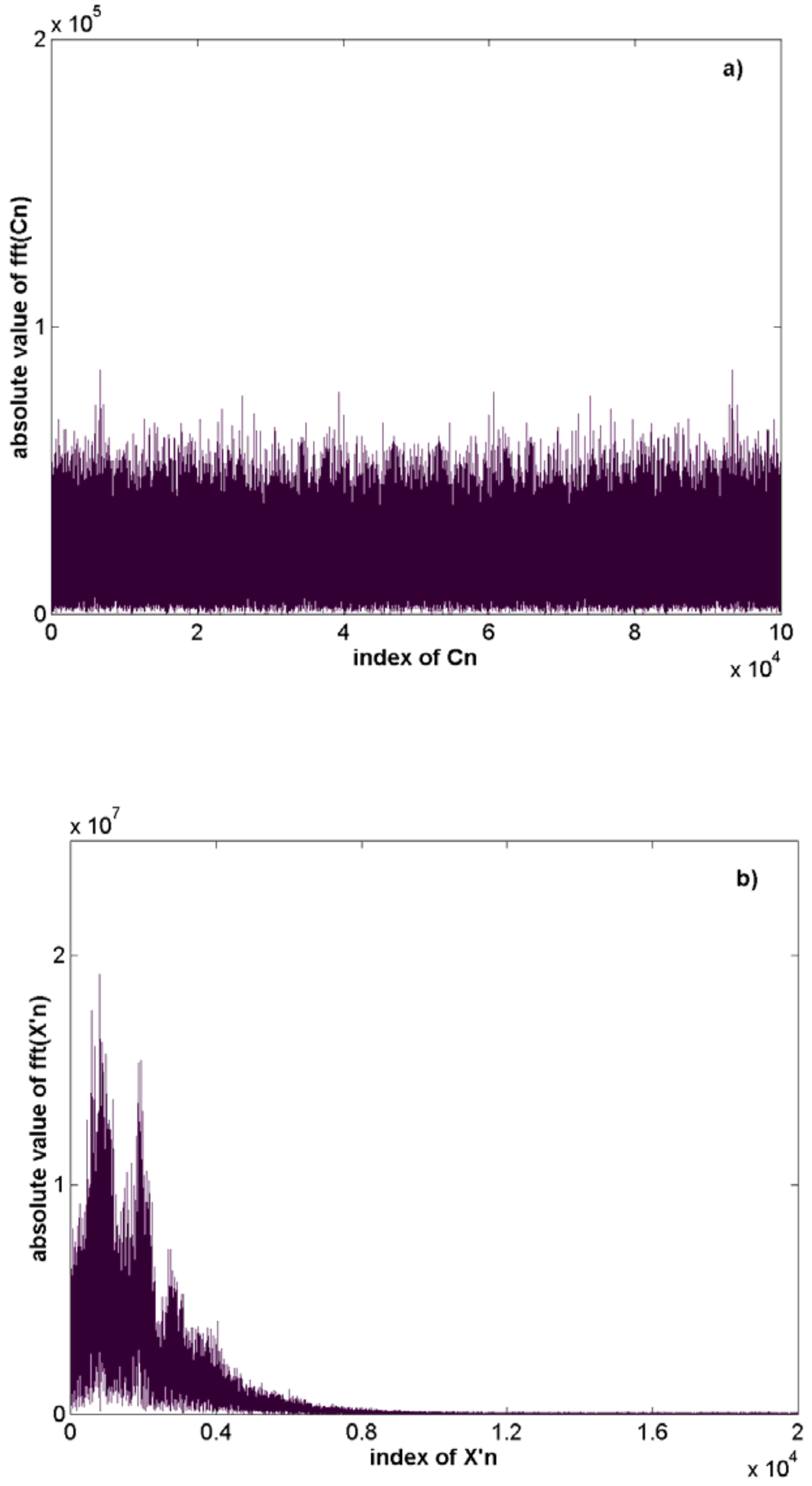

**Fig. 6.** Fast Fourier Transform (FFT) of the ciphered text $C_n$. The plain text is a periodic sequence of the numbers 0 to 255. The FFT is evaluated over 100,000 samples. (a) only the least significant byte of $X_n$ is used in the encryption process. (b) all the 17 bits of $X_n$ are used in the encryption process

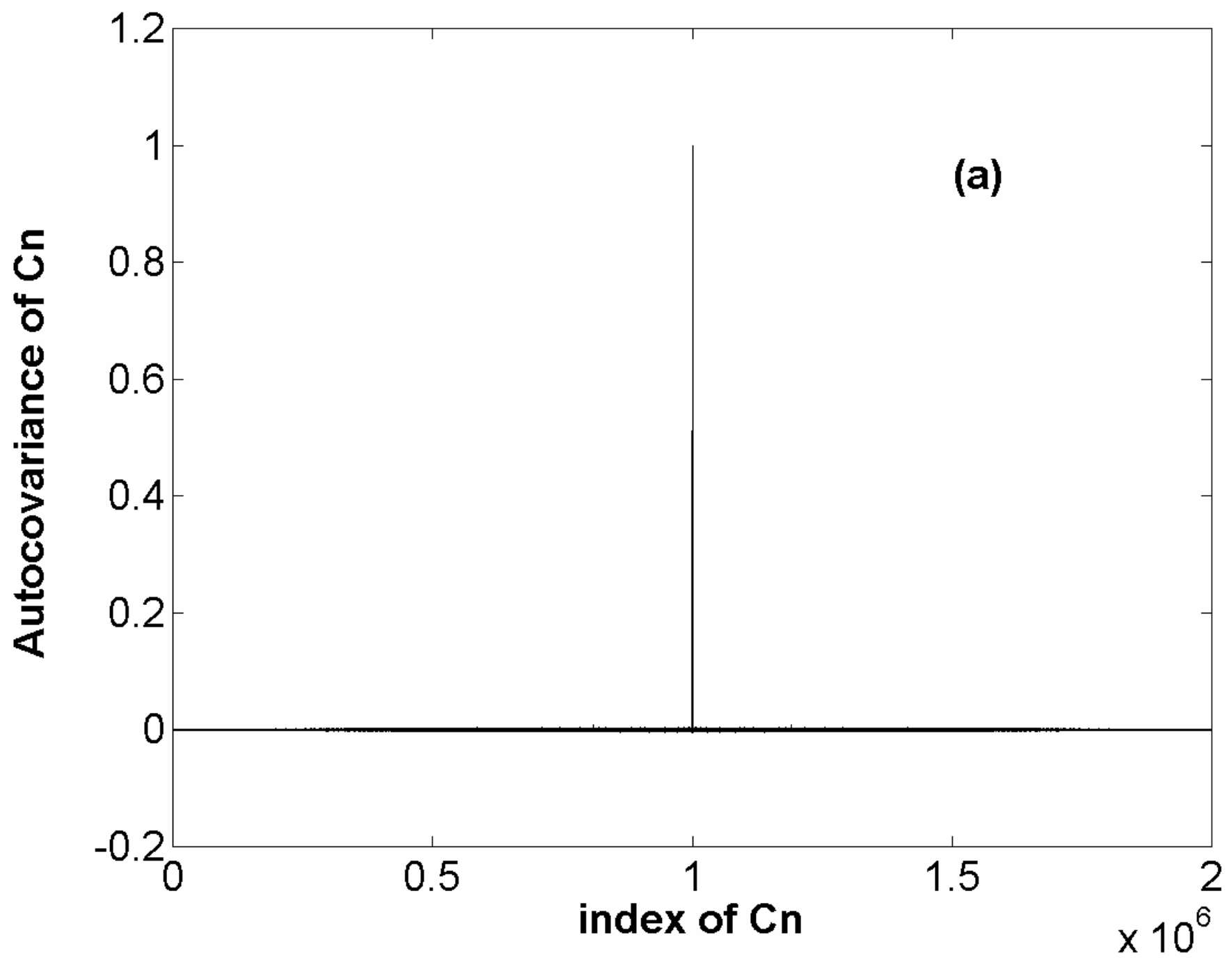

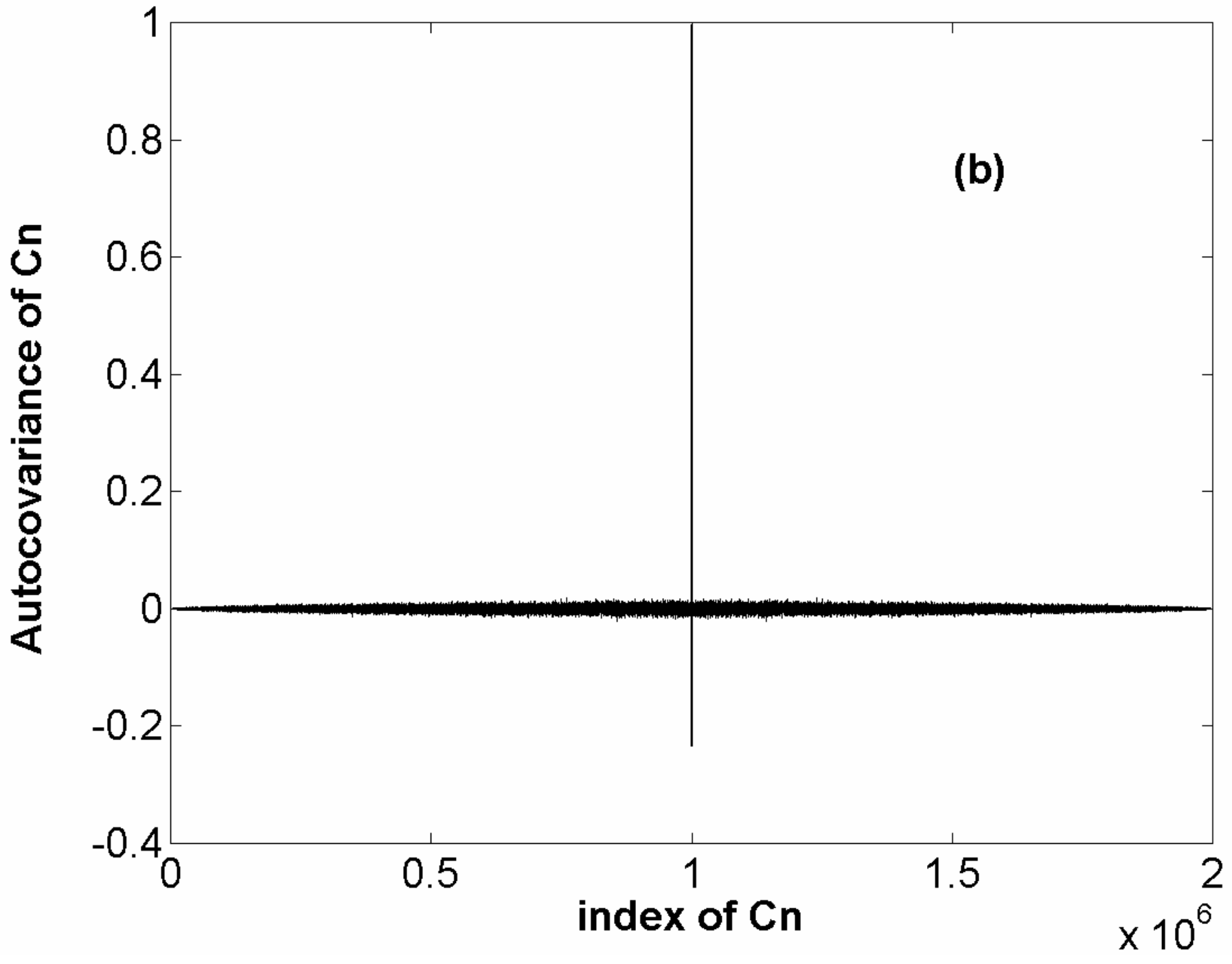

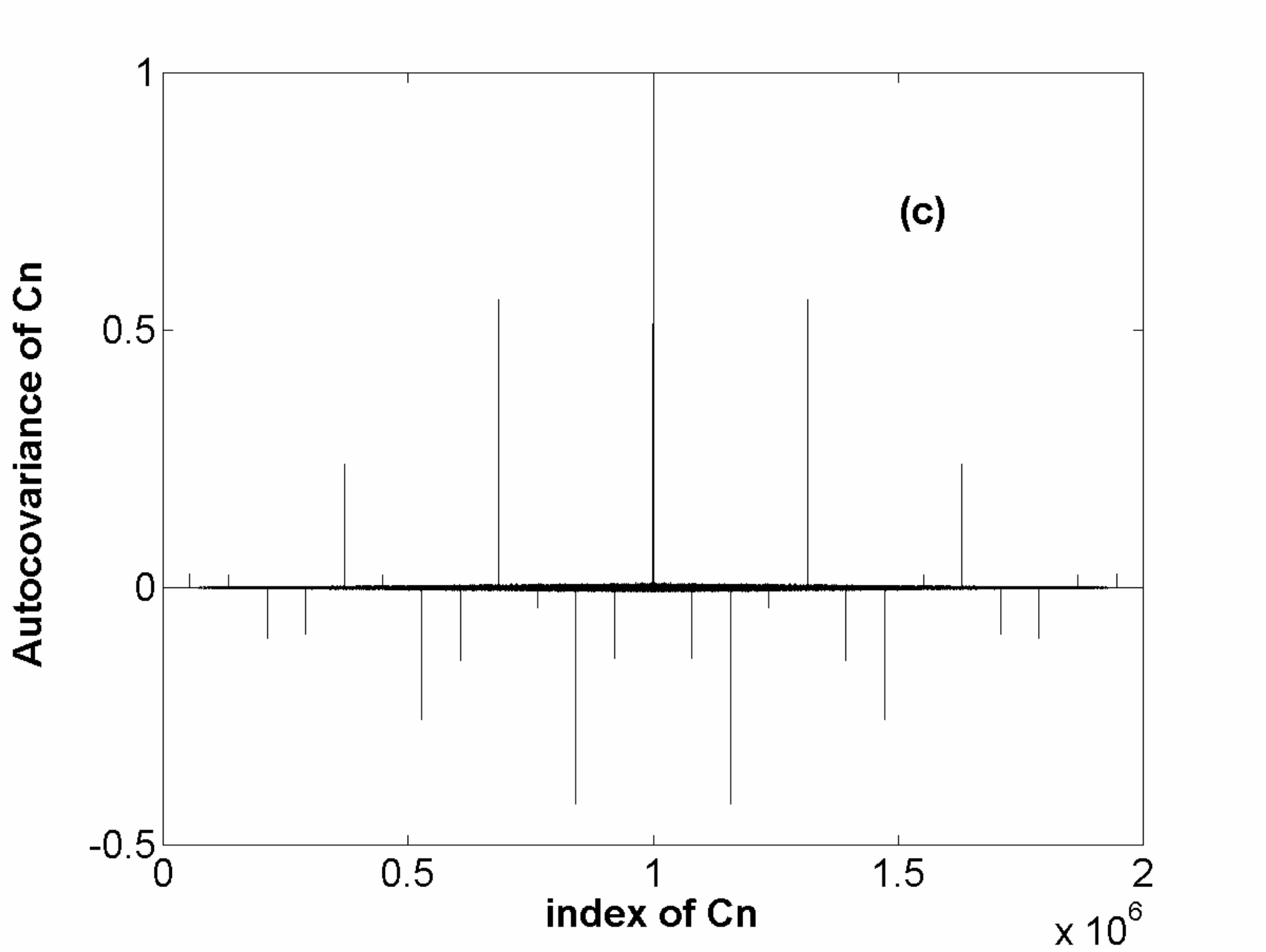

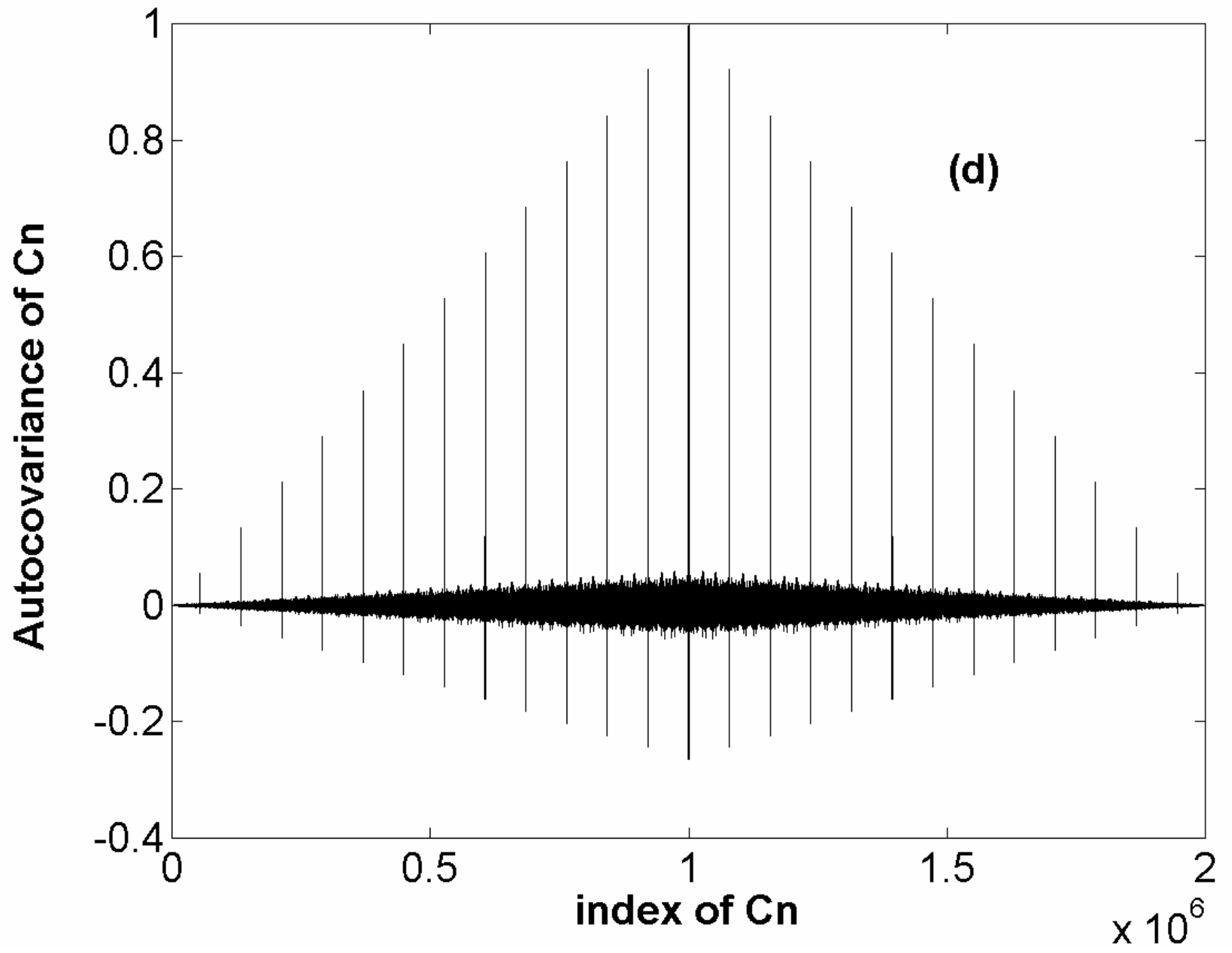

.**Fig. 7.** Autocovariance of the ciphered text $C_n$ evaluated over 1,000,000 samples: (a) only the least significant byte of $X_n$ is used in the encryption process and perturbation is on; (b) only the least significant byte of $X_n$ is used in the encryption process and perturbation is off; (c) all the 17 bits of $X_n$ are used in the encryption process and perturbation is on; (d) all the 17 bits of $X_n$ are used in the encryption process and perturbation is off.

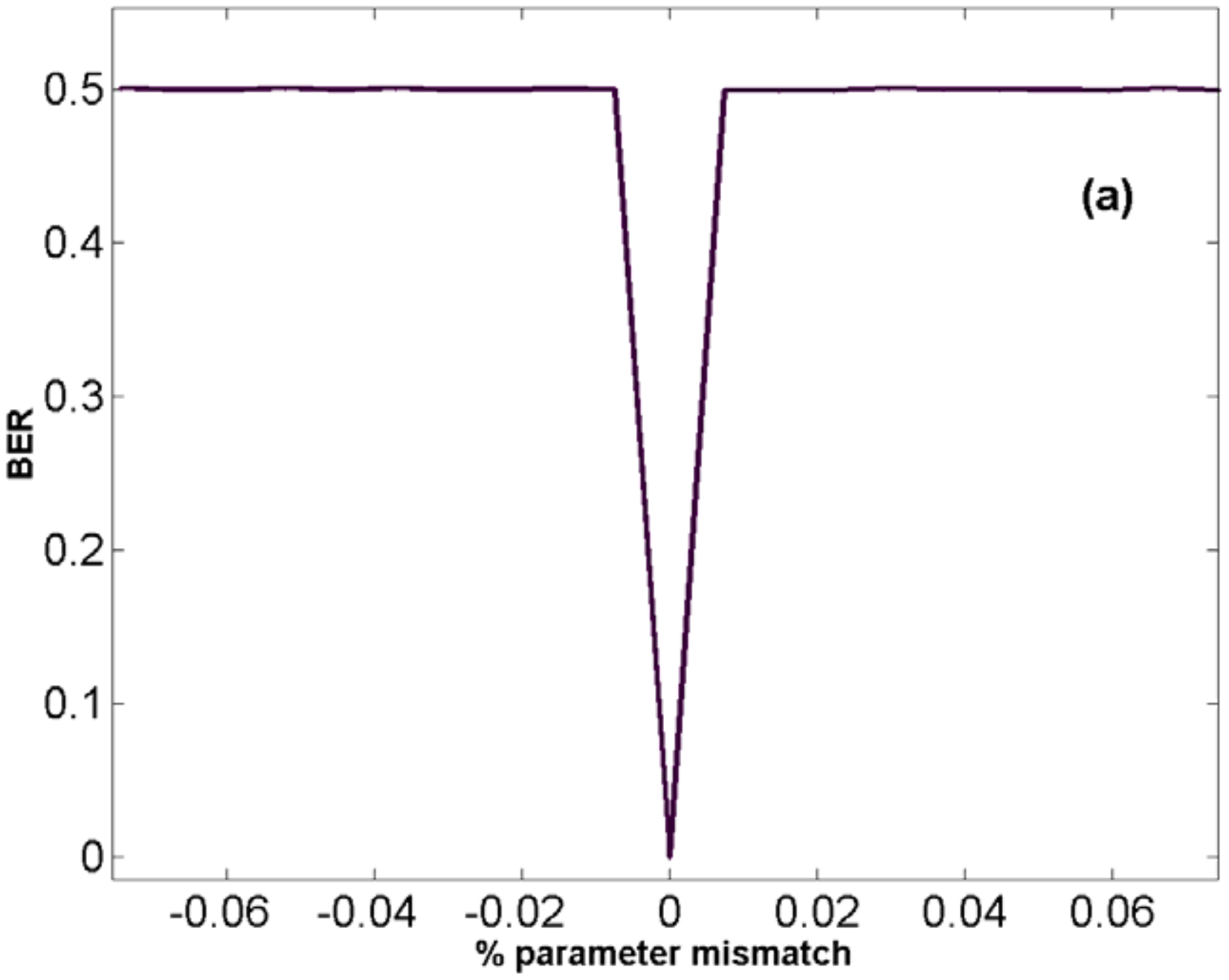

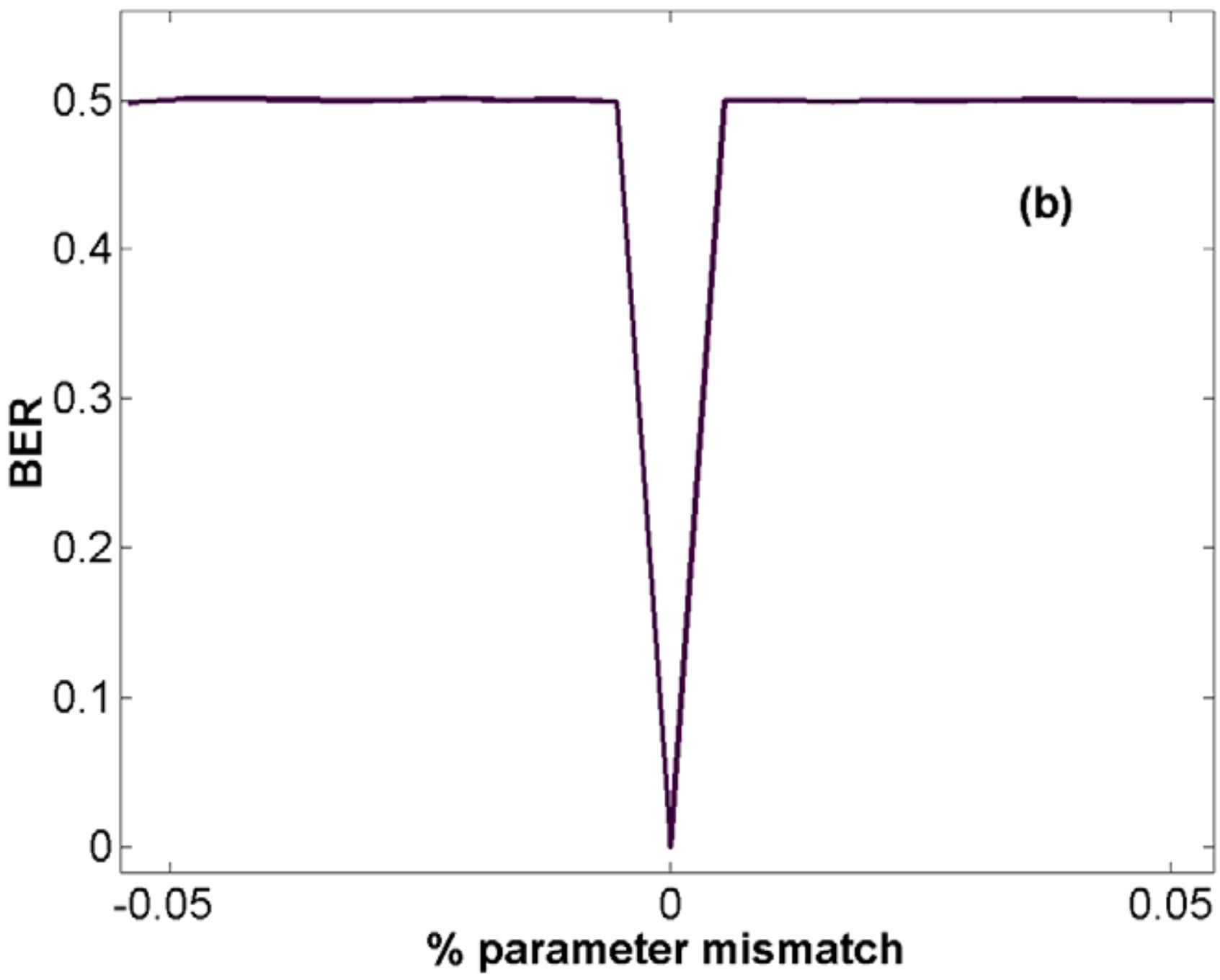

**Fig. 8.** Relationship between BER and percentage of receiver mismatches. (a) BER versus parameter $p$=($kB^2S$+$kbBS$) of equation 4. (b) BER versus a mismatch in the initial condition $x_0$

Table I
Comparison between different pseudo random generators using standard tests[23, 24]. The column labeled Lorenz correspond to the cipher presented in this paper. The other columns correspond to other pseudo random generators available in the literature. Each generator is tested five times. A shaded value indicates that the test has failed

| **Test** | **Generator** | | | |
|---|---|---|---|---|
| | **Lorenz** | **LFSR** | **Lehmer** | **Marsaglia** |
| **Frequency**<br>$n = 400{,}000$,<br>$\alpha = 0.05$, $\nu = 1$<br>If $X < 3.842$<br>=> accept $H_o$ | 1.875<br>0.031<br>0.161<br>0.038<br>1.823 | 1.170<br>2.134<br>1.513<br>0.005<br>0.188 | 0.313<br>0.083<br>0.015<br>3.576<br>0.520 | 1.239<br>2.510<br>0.062<br>0.013<br>1.399 |
| **Serial**<br>$n = 400{,}000$,<br>$\alpha = 0.05$, $\nu = 2$<br>If $X < 5.992$<br>=> accept $H_o$ | 1.960<br>0.329<br>1.546<br>1.157<br>3.533 | 4.721<br>2.136<br>2.821<br>0.158<br>2.242 | 0.643<br>0.860<br>2.882<br>3.658<br>1.490 | 1.35<br>2.512<br>0.137<br>2.468<br>1.427 |
| **Poker**<br>$n = 400{,}000$, (*)<br>$\alpha = 0.05$, $\nu = 255$<br>If $X < 293.248$<br>=> accept $H_o$ | 242.836<br>255.483<br>234.931<br>208.942<br>291.538 | 280.735<br>336.379<br>228.910<br>219.387<br>237.194 | 245.581<br>232.586<br>273.690<br>282.209<br>259.323 | 277.140<br>228.357<br>229.903<br>251.233<br>261.402 |
| **Runs**<br>$n = 400{,}000$,<br>$\alpha = 0.05$, $\nu = 26$<br>If $X < 38.885$<br>=> accept $H_o$ | 33.808<br>26.631<br>19.686<br>35.118<br>25.467 | 68.345<br>52.351<br>52.528<br>35.207<br>71.248 | 31.890<br>33.578<br>26.042<br>21.584<br>27.997 | 20.144<br>28.897<br>18.645<br>37.228<br>24.702 |
| **Autocorrelation**<br>$n = 400{,}000$,<br>$\alpha = 0.05$<br>If $\|X\| < 1.645$<br>=> accept $H_o$ | -0.293<br>-0.545<br>1.178<br>-1.058<br>1.304 | 1.886<br>0.043<br>1.140<br>0.391<br>-1.434 | 0.577<br>0.881<br>-1.693<br>-0.286<br>0.985 | 0.340<br>0.049<br>-0.270<br>1.567<br>0.169 |

(*) The sequence with length $n$ (400,000) is split into $k$ (50,000) non-overlapping parts each of length $m$ (8).